\documentclass[11pt]{article}
 
\usepackage[dvips]{graphicx}
\usepackage{geometry}
\usepackage{amsmath}
\usepackage{amssymb}
\usepackage{bbm}
\makeatletter
\@addtoreset{equation}{section}

\def\Hp{{\cal H_{\rm phys}}}
\def\d{\partial}
\def\sq{\hbox{\rlap{$\sqcap$}$\sqcup$}}
\def\eh{\scriptstyle{\frac{1}{2}}}
\def\fii{\varphi}
\def\al{\alpha}
\def\be{\beta}
\def\ro{\varrho}
\def\si{\sigma}
\def\pdh{(2\pi )^{-3/2}}
\def\io{\int\frac{d^3k}{\sqrt{2\omega}}\,}

\def\eps{\varepsilon}

\begin{document}

\title{QUANTUM GRAVITATIONAL BREMSSTRAHLUNG -- MASSLESS VERSUS
MASSIVE GRAVITY}
\author{Julian B. Berchtold \footnote{e-mail: julian@iisz.ch}
and G\"unter Scharf 
\footnote{e-mail: scharf@physik.unizh.ch}
\\ Institut f\"ur Theoretische Physik, 
\\ Universit\"at Z\"urich, 
\\ Winterthurerstr. 190 , CH-8057 Z\"urich, Switzerland
\\ and 
\\ Institute for Independent Studies Zurich
\\ Fortunagasse 18, CH-8001 Z\"urich, Switzerland}

\date{}

\maketitle

\vskip 3cm

\begin{abstract} The massive spin-2 quantum gauge theory previously
developed is applied to calculate gravitational bremsstrahlung. It is
shown that this theory is unique and free from defects. In particular, there is no
strong coupling if the graviton mass becomes small. The cross sections
go over smoothly into the ones of the massless theory in the limit of
vanishing graviton mass. The massless cross sections are calculated for
the full tensor theory.

\end{abstract}

\newpage

\section{Introduction and Summary}

Gravity with massive gravitons got much interest in recent years in view
of the evidence for dark energy. However, various massive gravity
theories suffer from serious defects. There may be a discontinuity if
the graviton mass $m_0$ goes to zero (VDVZ discontinuity), a violation of
Lorentz invariance or fields with wrong sign of the kinetic term in the
Lagrangian (\cite{VDR,AUB} and references given therein). It is misleading to call
those fields "ghosts"
because this name is occupied by the Fermi fields with integer spin
which are essential in gauge theories (Faddeev - Popov ghosts). The
reason for the failure is that those theories have been set up {\it without taking the gauge structure into account}. The early work by Boulware and Deser \cite{BD} can be criticized for the same reason. One
should never forget that gravity is a gauge theory side by side with
non-abelian gauge theories \cite{GHOST}. This is true for both, massless and
massive gravity. In fact, D.R.Grigore and one of us (G.S.) have
constructed the massive spin-2 gauge theory on Minkowski background according to the laws of
perturbative quantum gauge invariance \cite{GS}. The result was that this theory
is essentially unique and its classical limit agrees (in the pure
graviton sector) with general relativity with a cosmological constant
$\Lambda=m_0^2/2$. The consistency has been verified beyond linear gravity up to the quartic couplings of the graviton. A proof to all orders has been given in the massless case \cite{D}. There is no doubt that massive gravity considered as spin-2 quantum gauge theory is mathematically consistent. There remains the question whether it is also consistent with nature for the very small graviton mass $m_0=\sqrt{2\Lambda}$.

It has been shown in \cite{GS} that this theory has no discontinuity for
$m_0\to 0$ as far as the graviton propagator is concerned. Of course it
is fully Lorentz invariant. Another question is whether the
additional degrees of freedom of the massive graviton give rise to a low
strong coupling scale as in the theories (\cite{AUB}) mentioned above. This question is
studied here for gravitational bremsstrahlung because this process might
be close to experimental observations.

The origin of the so-called strong coupling is the longitudinal graviton
mode of the form (see \cite{AUB} eq.(6))
\begin{equation}
e_{\mu\nu}\sim\frac{k_\mu k_\nu}{m_0^2}+\ldots
\end{equation}
Since $m_0=\sqrt{2\Lambda}$ is very small, the scattering amplitudes
involving this mode get large. But this disaster is not real because
the longitudinal (massive) graviton states do not belong to the physical
Hilbert space $\Hp$. As well known in a gauge theory, $\Hp$ is defined by
means of the nilpotent gauge charge $Q$ ($Q^2=0$) in the following, two
possible forms \cite{GHOST}
\begin{eqnarray}
\Hp&=&{\rm Ker} Q/{\rm Ran}Q\\
\Hp&=&{\rm Ker}(QQ^++Q^+Q)\label{kerQ}
\end{eqnarray}
The second representation \ref{kerQ} is best suited for the construction of
the physical states in momentum space. For massive gravity, $Q$ is given
in terms of the asymptotic free quantum fields by
\begin{equation}\label{Q}
Q\stackrel{\rm def}{=} \int\limits_{x^0=t} d^3x\,\Bigl[\d_\nu h^{\mu\nu}(x)+m_0 
v^\mu(x)\Bigl]\overleftrightarrow{\d}_0\,u_\mu(x)
\end{equation}
Here $h^{\mu\nu}$ is the symmetric tensor field, $v^\mu$ a Bose vector
field and $u_\mu$ is the fermiomic ghost field. All fields satisfy
Klein-Gordon equations with mass $m_0$, further properties are discussed
in the next section.

To study massive gravity in the neighborhood of $m_0=0$, we choose a
Lorentz frame where the graviton momentum is $k_\mu=(\omega,0,0,k_3)$,
$\omega^2=k_3^2+m_0^2$. Then in section 3, we determine the kernel of the
selfadjoint operator in \ref{kerQ} by calculating its eigenfunctions for
eigenvalue 0. The kernel consists of 6 modes. Two modes agree exactly with the two transverse graviton states of
the massless theory. The other four modes have contributions from the
emission operators of the vector field $v^\mu$. In the limit $m_0\to 0$,
only the latter survive. For this reason, we call $v^\mu$ the
vector-graviton field. Note that no longitudinal mode contributes to the
physical subspace in this representation. Since the vector-graviton
does not couple to ordinary matter in a gauge invariant way, the
cross-sections for bremsstrahlung go over smoothly into the massless ones
in the limit $m_0\to 0$.

In the last section, we calculate the differential cross-section and the
total radiated energy of quantum gravitational bremsstrahlung in the
lowest order of perturbation theory for $m_0=0$. It seems that this has
not been done before for the full tensor theory. In particular, we
consider the case of an ultrarelativistic particle being scattered by a
heavy mass $M$ through a small angle. Under these assumptions, our
results agree with classical calculations.

\section{Field content and gauge structure}

The basic free asymptotic fields of massive gravity are the symmetric
tensor field $h^{\mu\nu}(x)$ with arbitrary trace, the fermionic ghost
$u^\mu(x)$ and anti-ghost $\tilde u^\mu(x)$ fields and the bosonic
vector-graviton field $v^\mu(x)$. They all satisfy the massive
Klein-Gordon eqations
\begin{equation}
(\sq+m_0^2)h^{\mu\nu}=0=(\sq+m_0^2)u^\mu\ldots
\end{equation}
and are quantized as follows \cite{GS}
\begin{equation}\label{hquant}
\big[h^{\alpha\beta}(x), h^{\mu\nu}(y)\big]=-ib^{\alpha\beta\mu\nu}D_{m_0}(x-y)
\end{equation}
with
\begin{eqnarray}
b^{\alpha\beta\mu\nu}&=&{\scriptstyle{\frac{1}{2}}}\big(\eta^{\alpha\mu}\eta^{\beta\nu}+\eta^
{\alpha\nu}\eta^{\beta\mu}-\eta^{\alpha\beta}\eta^{\mu\nu}\big)\label{b-tensor}\\
\big\{u^\mu(x),\tilde u^\nu(y)\big\}&=&i\eta^{\mu\nu}D_{m_0}(x-y)\\
\big[v^\mu(x),\, v^\nu(y)\big]&=&{\scriptstyle{\frac{1}{2}}}\eta^{\mu\nu}D_{m_0}(x-y)\label{comvg}
\end{eqnarray}
and zero otherwise. Here, $D_{m_0}(x)$ is the Jordan-Pauli distribution 
with mass $m_0$ and $\eta^{\mu\nu}={\rm diag}(1,-1,-1,-1)$ the Minkowski tensor.
Note that the vector-graviton field $v^\mu$ is essential to render the
gauge charge $Q$ nilpotent:
\begin{eqnarray}
Q^2\;\;=\;\;{\eh}\{Q,Q\}&=&\frac{1}{2}\int d^3x(\d_\nu h^{\mu\nu}+mv^\mu)\{\overleftrightarrow{\d}_0\,u_\mu,Q\}\nonumber\\
& &-\frac{1}{2}\int d^3x[\d_\nu h^{\mu\nu}+mv^\mu,Q]\overleftrightarrow{\d}_0\,u_\mu\nonumber\\
&=&0\nonumber
\end{eqnarray}
The factor 1/2 in \ref{comvg} is due to the convention \ref{b-tensor}.

The commutators or anticommutators with $Q$, respectively, define the
gauge variations of the asymptotic fields. The commutation relations 
immediately yield
\begin{eqnarray}
d_Q h^{\mu\nu}\;\;\stackrel{\rm def}{=}\;\;[Q,h^{\mu\nu}]&=&-{\scriptstyle{\frac{i}{2}}}\big(\d^\nu u^\mu+\d^\mu u^\nu-\eta^{\mu\nu}\d_\al u^\al\big)\label{dQh}\\
d_Q u^\mu&\stackrel{\rm def}{=}&\big\{Q,u\big\}=0\\
d_Q\tilde u^\mu&\stackrel{\rm def}{=}&\big\{Q,\tilde u^\mu\big\}\;\; =\;\;i\big(\d_\nu h^{\mu\nu}-m_0 v^\mu\big)\label{dQut}\\
d_Q v^\mu&\stackrel{\rm def}{=}&\big[Q, v^\mu\big]\;\;=\;\;-{\scriptstyle{\frac{i}{2}}}m_0u^\mu\label{dQv}
\end{eqnarray}
Note that if \ref{dQh} is considered as an equation between classical fields,
this is just the infinitesimal version of the diffeomorphic coordinate
transformations of general relativity. 

Now we are able to formulate gauge invariance of the S-matrix. Let
\begin{equation}\label{Smatrix}
S(g)=\sum_{n=0}^\infty\frac{1}{n!}\int dx_1\cdots dx_n\,
T_n(x_1,\ldots,x_n)\,g(x_1)\cdots g(x_n)
\end{equation}
be the perturbative S-matrix where $g(x)$ is a Schwartz test function
which switches the interaction. The time-ordered products $T_n$ are
expressed in terms of free fields defined above. Then $d_Q T_n$ is well
defined and gauge invariance of $S(g)$ is defined perturbatively as
follows: 

first order gauge invariance:
\begin{equation}\label{dQT1}
d_Q T_1(x)=i\frac{\d}{\d x^\mu}T_{1/1}^\mu(x)
\end{equation}

$n$-th order gauge invariance:
\begin{equation}\label{dQTn}
d_QT_n=i\sum_{l=1}^n\frac{\d}{\d x_l^\mu}T\big\{T_1(x_1)\ldots T_{1/1}^\mu 
(x_l)\cdot\ldots T_1(x_n)\big\}
\end{equation}
where the time-ordered products must be appropriately normalized. It is
a very nice feature that the conditions \ref{dQT1} and \ref{dQTn} for $n=2$
determine the theory essentially uniquely (that means up to divergence
and coboundary couplings). No classical Lagrangian is needed. It was
shown in \cite{GS} that in case of the massive spin-2 theory we have:
\begin{eqnarray}\label{T1}
T_1&=&h^{\mu\nu}\d_\mu h\d_\nu h-2h^{\mu\nu}\d_\mu h_{\al\be}\d_\nu
h^{\al\be}-4h_{\mu\nu}\d_\al h^{\mu\be}\d_\be h^{\nu\al}\nonumber\\
& &-2h_{\mu\nu}\d_\al h^{\mu\nu}\d^\al h+4h_{\mu\nu}\d_\al h_{\mu\be}
\d^\al h_\nu^{\,\be}\nonumber\\
& &+4\d_\mu h^{\mu\nu}u^\al\d_\al\tilde u_\nu-4h^{\mu\nu}\d_\mu u^\al
\d_\nu\tilde u_\al+4h^{\mu\nu}\d_\mu v^\al\d_\nu v_\al\nonumber\\
& &-4m_0\d_\mu v_\nu u^\mu\tilde u^\nu+m_0^2\Big(-\frac{4}{3}h^{\mu\nu}
h_{\mu\al}h_\nu^{\,\al}+h^{\mu\nu}h_{\mu\nu}h-\frac{1}{6}h^3\Big)
\end{eqnarray}
up to a coupling constant, where $h=h^\mu_{\,\mu}$ is the trace; all
products are normally ordered. As mentioned in the introduction, this result is in agreement with the expansion of the Einstein-Hilbert Lagrangian with a cosmological constant \cite{GS}.

For bremsstrahlung we need the gravity-matter couplings. We represent
the matter by a non-hermitian scalar field $\fii$ of mass $m$. Its coupling to 
$h_{\mu\nu}$ can be derived from gauge invariance as in the massless
case (\cite{GHOST}, sect.5.9):
\begin{equation}\label{coup}
T^m(x)=i\kappa\Big(h^{\mu\nu}\d_\mu\fii^+\d_\nu\fii-\frac{m^2}{2}h
\fii^+\fii\Big)
\end{equation}
But in the massive case, there may also be a coupling between $\fii$ and the
vector-graviton field $v^\mu$ of the form
\begin{equation}
T_1=a_1\d_\mu v^\mu\fii^+\fii+a_2v^\mu\d_\mu\fii^+\fii+a_3v^\mu
\fii^+\fii
\end{equation}
In order to fulfill first order gauge invariance \ref{dQT1}, $d_QT_1$ must be
a divergence. Using $d_Q\fii=0$ and \ref{dQv} we obtain
\begin{equation}\nonumber
d_QT_1=-\frac{i}{2}m_0a_1\d_\mu v^\mu\fii^+\fii-\frac{i}{2}m_0(a_2u^\mu
\d_\mu\fii^+\fii+a_3u^\mu\fii^+\d_\mu\fii)
\end{equation}
This must be a divergence
\begin{equation}\nonumber
=-\frac{i}{2}m_0\d_\mu(b_1u^\mu\fii^+\fii)=-\frac{i}{2}b_1(\d_\mu u^\mu
\fii^+\fii+u^\mu\d_\nu\fii^+\fii+u^\mu\fii^+\d_\mu\fii)
\end{equation}
Then it follows that $b_1=a_1=a_2=a_3$ and
\begin{equation}\nonumber
T_1=a_1\d_\mu(v^\mu\fii^+\fii)
\end{equation}
Since such a divergence coupling is physically irrelevant, there exists
no non-trivial gauge invariant coupling between $v^\mu$ and $\fii$.

Finally, the interaction of the $\fii$-particle with an external
classical potential is described by \ref{coup} where $h^{\mu\nu}$
is substituted by $h^{\mu\nu}_{\rm ext}$. In the Newtonian limit, which
we will consider later, we have
\begin{equation}
h^{00}=-2\phi_N
\end{equation}
where $\phi_N$ is the Newtonian potential.

\section{The physical Hilbert space for massive gravitons}

To get a concrete Hilbert space representation, we must express the
various fields by means of emission and absorption operators. We follow
the discussion of the massless case in \cite{GHOST} as close as possible. We
decompose $h^{\al\be}$ into its traceless part and the trace $h$
\begin{equation}
h^{\alpha\beta}(x)=H^{\alpha\beta}(x)+\frac{1}{4}\eta^{\alpha\beta}
h(x)
\end{equation}
From \ref{hquant} we obtain the following commutation relations
\begin{eqnarray}
\big[h(x),h(y)\big]&=&4iD_{m_0}(x-y)\label{hcom}\\
\big[H^{\alpha\beta}(x), H^{\mu\nu}(y)\big]&=&-it^{\alpha\beta\mu\nu}D_{m_0}(x-y)\label{Hcom}
\end{eqnarray}
with
\begin{equation}
t^{\alpha\beta\mu\nu}\stackrel{\rm def}{=}{\eh}\big(\eta^{\alpha\mu}\eta^{\beta\nu}+\eta^{\alpha\nu} 
\eta^{\beta\mu}-{\eh}\eta^{\alpha\beta}\eta^{\mu\nu}\big)=t^{\mu\nu\alpha\beta}
\end{equation}
and
\begin{equation}
\big[H^{\alpha\beta}(x), h(y)\big]=0
\end{equation}
We claim that the fields in \ref{hcom} and \ref{Hcom} can be represented as follows:
\begin{equation}
H^{\alpha\beta}(x)=\pdh\io\,\Bigl( a_{\alpha\beta}(\vec k)e^{-ikx}
+\eta^{\alpha\alpha}\eta^{\beta\beta}a_{\alpha\beta}^+(\vec k)e^{ikx}\Bigl)
\end{equation}
where $a_{\alpha\beta}=a_{\beta\alpha}$ is symmetric and satisfies the
commutation relation
\begin{equation}\label{tcom}
\big[a_{\alpha\beta}(\vec k), a_{\mu\nu}^+(\vec k')\big]=\eta^{\alpha\alpha}
\eta^{\beta\beta}t^{\alpha\beta\mu\nu}\delta(\vec k-\vec k')
\end{equation}
The trace part is given by
\begin{equation}
h(x)=\pdh\io\Bigl(d(\vec k)e^{-ikx}-d^+(\vec k)e^{ikx}\Bigl)
\end{equation}
with
\begin{equation}\label{dcom}
\big[d(\vec k), d^+(\vec k')\big]=4\delta(\vec k-\vec k')
\end{equation}
Since the right-hand side is positive, the $h$-sector of the Fock space can be
constructed in the usual way by applying products of $d^+$'s to the
vacuum.

The situation is not so simple in the $H$-sector because the righthand
side of \ref{tcom} is not a diagonal matrix. 
We perform a linear transformation of the diagonal
operators $a_{\alpha\alpha}$ and $a_{\alpha\alpha}^+$
in such a way that the new operators are usual annihilation and creation
operators
\begin{equation}\label{acom}
\big[\tilde a_{\alpha\alpha}(\vec k),\tilde a_{\beta\beta}^+(\vec k')\big]
=\delta_{\alpha\beta}\delta(\vec k-\vec k')
\end{equation}
The following transformation does the job:
\begin{eqnarray}
a_{00}&=&{\eh}\big(\tilde a_{11}+\tilde a_{22}+\tilde a_{33}\big)\nonumber\\
a_{11}&=&{\eh}\big(-\tilde a_{11}+\tilde a_{22}+\tilde a_{33}\big)\nonumber\\
a_{22}&=&{\eh}\big(\tilde a_{11}-\tilde a_{22}+\tilde a_{33}\big)\nonumber\\
a_{33}&=&{\eh}\big(\tilde a_{11}+\tilde a_{22}-\tilde a_{33}\big)\label{transf}
\end{eqnarray}
We note that $\tilde a_{00}$ does not appear because one pair of absorption
and emission operators is superfluous due to the trace condition
$H^\alpha\,_\alpha=0$. In fact, from \ref{transf} we see
\begin{equation}
\sum_{j=1}^3a_{jj}=a_{00}
\end{equation}
The Fock representation can now be constructed as usual by means of
$\tilde a_{11}^+,\tilde a_{22}^+,\tilde a_{33}^+$ and $a_{\alpha\beta}
^+$ with $\alpha\ne\beta$.

The other fields have the following representation in terms of emission
and absorption operators:
\begin{eqnarray}
u^\mu(x)&=&\pdh\int\frac{d^3k}{\sqrt{2E_k}}\,\Bigl(c_2^\mu(\vec
k)e^{-ipx}-\eta^{\mu\mu}c_1^\mu(\vec k)^+e^{ipx}\Bigl)\\
\tilde u^\mu(x)&=&\pdh\int\frac{d^3k}{\sqrt{2E_k}}\,
\Bigl(-c_1^\mu(\vec k)e^{-ipx}-\eta^{\mu\mu}c_2^\mu
(\vec k)^+e^{ipx}\Bigl)\\
v^\mu(x)&=&\pdh\int\frac{d^3k}{2\sqrt{E_k}}\,\Bigl(b^\mu(\vec
k)e^{-ipx}-\eta^{\mu\mu}b^\mu(\vec k)^+e^{ipx}\Bigl)
\end{eqnarray}
with the following (anti)commutation relations
\begin{eqnarray}
\big\{c_j^\mu(\vec k),c_l^\nu(\vec k')^+\big\}&=&\delta_{jl}\delta^\mu_{\nu} 
\delta^3(\vec k-\vec k')\\
\big[b^\mu(\vec k), b^\nu(\vec k)^+\big]&=&\delta^\mu_\nu\delta^3(\vec k-\vec
k')\label{bcom}
\end{eqnarray}
Then the gauge charge $Q$ (\ref{Q}) can be written in momentum space as
follows 
\begin{equation}
Q=\int d^3k\,\Bigl(A^\alpha(\vec k)^+c_2^\gamma(\vec k)-B^\alpha(\vec k)
c_1^\gamma(\vec k)^+\Bigl)\eta_{\alpha\gamma}
\end{equation}
where
\begin{eqnarray}
A^\alpha&=&\eta^{\alpha\alpha}a^{\alpha\beta}(\vec k)k^\beta-
\frac{k^\alpha}{4}d(\vec k)-im_1\eta^{\al\al}b^\al\\
B^\alpha&=&\big(a^{\alpha\beta}(\vec k)k_\beta+
\frac{k^\alpha}{4}d(\vec k)+im_1b^\al\big)\eta^{\alpha\alpha}
\end{eqnarray}
The adjoint is given by
\begin{equation}
Q^+=\int d^3k\,\Bigl(c_2^\delta(\vec k)^+A^\be(\vec k)-
c_1^\delta(\vec k)B^\be(\vec k)^+\Bigl)\eta_{\delta\be}
\end{equation}
where
\begin{equation}
m_1=\frac{m_0}{\sqrt{2}}
\end{equation}
According to \ref{kerQ}, the physical Hilbert space is the kernel of the
selfadjoint operator
\begin{eqnarray}
\big\{Q,Q^+\big\}&=&\int d^3k\,d^3k'\,\Big(A^\alpha(\vec k)^+A^\beta(\vec k')
\big\{c_2^\gamma(\vec k),c_2^\delta(\vec k')^+\big\}\nonumber\\
& &+B^\beta(\vec k')^+B^\alpha(\vec k)\big\{c_1^\delta(\vec k'),c_1^\gamma
(\vec k)^+\big\}+c_2^\delta(\vec k')^+c_2^\gamma(\vec k)\big[A^\beta(\vec k'), 
A^\alpha(\vec k)^+\big]\nonumber\\
& &+c_1^\gamma(\vec k)^+c_1^\delta(\vec k')\big[B^\alpha(\vec k),
B^\beta(\vec k')^+\big]\Big)\eta_{\alpha\gamma}\eta_{\beta\delta}
\end{eqnarray}
We restrict to the graviton sector because the ghost sector is
unphysical.
\begin{equation}\label{QQ}
\big\{ Q,Q^+\big\}\vert_{\rm graviton}
=\int d^3k\,\sum_{\alpha=0}^3\Big(A^{\alpha +}A^\alpha+
B^{\alpha +}B^\alpha\Big)
\end{equation}
It is convenient to introduce time-like and space-like components:
\begin{eqnarray}
A^0&=&k_0\big(a^{00}-a^0_\parallel-\frac{d}{4}-\frac{im_1}{k_0}b^0\big)\nonumber\\
A^j&=&k_0\big(-a^{0j}+a^j_\parallel-\frac{k^j}{k_0}\frac{d}{4}+\frac{im_1}{k_0}b^j\big)\nonumber\\
B^0&=&k_0\big(a^{00}+a^0_\parallel+\frac{d}{4}+\frac{im_1}{k_0}b^0\big)\nonumber\\
B^j&=&k_0\big(-a^{0j}-a^j_\parallel-\frac{k^j}{k_0}\frac{d}{4}-\frac{im_1}{k_0}b^j\big)
\end{eqnarray}
where
\begin{equation}
a_\parallel^\mu=\frac{k_j}{k_0}a^{\mu j}
\end{equation}
We choose a Lorentz frame where $k_\mu=(\omega,0,0,k_3)$. Then
we get for the integrand in \ref{QQ}
\begin{eqnarray}\label{QF}
\sum^3_{\alpha=0}\Big(A^{\alpha+}A^\alpha+B^{\alpha+}B^\alpha\Big)&=&2\omega^2\,\bigg\{a^{00+}a^{00}+\frac{k^2_3}{\omega^2}\,a^{03+}a^{03}\nonumber\\
& &\;\;\;\;\;\;\;+a^{01+}a^{01}+a^{02+}a^{02}+a^{03+}a^{03}\nonumber\\
& &\;\;\;\;\;\;\;+\frac{k^2_3}{\omega^2}\,\Big[a^{13+}a^{13}+a^{23+}a^{23}+a^{33+}a^{33}\Big]\nonumber\\
& &\;\;\;\;\;\;\;+\frac{1}{16}\big(1+\frac{k_3^2}{\omega^2}\big)d^+d+\frac{im_1}{4\omega}\,d^+b^0-\frac{im_1}{4\omega}\,b^{0+}d\nonumber\\
& &\;\;\;\;\;\;\;+\frac{im_1k_3}{\omega^2}\,a^{03+}b^0-\frac{im_1k_3}{\omega^2}\,b^{0+}a^{03}\nonumber\\
& &\;\;\;\;\;\;\;+\frac{im_1k_3}{\omega^2}\,\Big[a^{13+}b^1+a^{23+}b^2+a^{33+}b^3\nonumber\\
& &\;\;\;\;\;\;\;\;\;\;-\,b^{1+}a^{13}-b^{2+}a^{23}-b^{3+}a^{33}\Big]\nonumber\\
& &\;\;\;\;\;\;\;+\frac{m_1^2}{\omega^2}\,\Big[b^{0+}b^0+\sum^3_{j=1}b^{j+}b^j\Big]\bigg\}
\end{eqnarray}
Since $a^{12+}$ does not appear herein, the states $a^{12+}\Omega$, where
$\Omega$ is the Fock vacuum, certainly belong to the kernel of \ref{QQ}
and, hence, are in the physical subspace.
We have still to substitute the diagonal operators $a^{\mu\mu}$ by means
of \ref{transf} by the operators $\tilde a^{jj}$ which generate the Fock states.
Then the quadratic form \ref{QF} can be represented in matrix notation
$A^+XA$ where $A^+$ stands for the emission operators
\begin{equation}
A^+=(\tilde a_{11}^+,\tilde a_{22}^+,\tilde a_{33}^+,\sqrt{2}a_{01}^+,\sqrt{2}a_{02}^+,
\sqrt{2}a_{03}^+,\sqrt{2}a_{13}^+,\sqrt{2}a_{23}^+,1/2d^+,b_0^+,b_1^+,b_2^+,b_3^+)
\end{equation}
where the numerical factors are necessary to get the states correctly normalized due to \ref{tcom}, \ref{dcom}, \ref{acom} and \ref{bcom}. According to \ref{QF}, $X$ is the hermitian $13\times 13$ matrix
\begin{eqnarray}
X&=&\left(
\begin{array}{ccccccc}
z&z&m_1^2&0&0&0&0\\
z&z&m_1^2&0&0&0&0\\
m_1^2&m_1^2&z&0&0&0&0\\
0&0&0&\omega^2&0&0&0\\
0&0&0&0&\omega^2&0&0\\
0&0&0&0&0&2z&0\\
0&0&0&0&0&0&k^2_3\\
0&0&0&0&0&0&0\\
0&0&0&0&0&0&0\\
0&0&0&0&0&-\sqrt{2}im_1k_3&0\\
0&0&0&0&0&0&-\sqrt{2}im_1k_3\\
0&0&0&0&0&0&0\\
-im_1k_3&-im_1k_3&im_1k_3&0&0&0&0
\end{array}\right.\nonumber\\
\;\nonumber\\
\;\nonumber\\
& &\left.
\begin{array}{cccccc}
0&0&0&0&0&im_1k_3\\
0&0&0&0&0&im_1k_3\\
0&0&0&0&0&-im_1k_3\\
0&0&0&0&0&0\\
0&0&0&0&0&0\\
0&0&\sqrt{2}im_1k_3&0&0&0\\
0&0&0&\sqrt{2}im_1k_3&0&0\\
k^2_3&0&0&0&\sqrt{2}im_1k_3&0\\
0&z&im_1\omega&0&0&0\\
0&-im_1\omega&2m_1^2&0&0&0\\
0&0&0&2m_1^2&0&0\\
-\sqrt{2}im_1k_3&0&0&0&2m_1^2&0\\
0&0&0&0&0&2m_1^2
\end{array}\label{X}
\right)
\end{eqnarray}
where
\begin{equation}
z={\eh}\big(\omega^2+k^2_3\big)
\end{equation}
The eigenvalues of this matrix are:
\begin{eqnarray}
\lambda_1&=&\lambda_2\;\;=\;\;\lambda_3\;\;=\;\;\lambda_4\;\;=\;\;\lambda_5\;\;=\;\;0\nonumber\\
\lambda_6&=&\lambda_7\;\;=\;\;\lambda_8\;\;=\;\;\lambda_9\;\;=\;\;\lambda_{10}\;\;=\;\;\lambda_{11}\;\;=\;\;\omega^2\;\;=\;\;2m^2_1+k^2_3\nonumber\\
\lambda_{12}&=&\lambda_{13}\;\;=\;\;3m^2_1+2k^2_3\nonumber\\
\end{eqnarray}
The 5 eigenvectors with eigenvalue 0 determine the kernel. They can be
easily calculated from the matrix \ref{X}. Together with the previously
found state $a_{12}^+\Omega$ we have 6 physical massive graviton modes:
\begin{eqnarray}
\psi_1&=&\sqrt{2}a_{12}^+\Omega\\
\psi_2&=&c_2\Bigl(\frac{im_1}{k_3}(\tilde a_{11}-\tilde a_{33})+b_3\Bigl)^+\Omega\\
& &c_2\;\;=\;\;\Bigl(1+\frac{2m_1^2}{k_3^2}\Bigl)^{-1/2}\nonumber\\
\psi_3&=&c_3\Bigl(\frac{2im_1}{k_3}\,a_{23}+b_2\Bigl)^+\Omega\\
& &c_3\;\;=\;\;\Bigl(1+\frac{2m_1^2}{k_3^2}\Bigl)^{-1/2}\nonumber\\
\psi_4&=&c_4\Bigl(\frac{2im_1}{k_3}\,a_{13}+b_1\Bigl)^+\Omega\\
& &c_4\;\;=\;\;\Bigl(1+\frac{2m_1^2}{k_3^2}\Bigl)^{-1/2}\nonumber\\
\psi_5&=&c_5\Bigl(\frac{im_1}{2(m_1^2+k_3^2)}\big(2k_3a_{03}+\sqrt{2m_1^2+k_3^2}\,d\big)+b_0
\Bigl)^+\Omega\\
& &c_5\;\;=\;\;\Bigl(1+m_1^2\,\frac{3k_3^2+4m_1^2}{2(m_1^2+k_3^2)^2}\Bigl)^{-1/2}\nonumber\\
\psi_6&=&\frac{1}{\sqrt{2}}(\tilde a_{11}-\tilde a_{22})^+\Omega
\end{eqnarray}
All physical states are normalized due to \ref{tcom}, \ref{dcom}, \ref{acom} and \ref{bcom}.

The states $\psi_1$ and $\psi_6$ agree exactly with the two transverse
physical modes of the massless graviton. The other four massive physical graviton
modes converge in the massless limit to the four vector-graviton states
$b_\mu^+\Omega$, $\mu=0,1,2,3$. We have seen in the last section that
the latter do not couple to matter in a gauge invariant way. This is the reason why the
scattering amplitudes for bremsstrahlung go over smoothly to the
massless ones in the limit $m_0\to 0$. We will return to this in
the next section.

\section{Quantum gravitational bremsstrahlung}

\begin{figure}[htb]
\begin{center}
\includegraphics[width=1\textwidth]{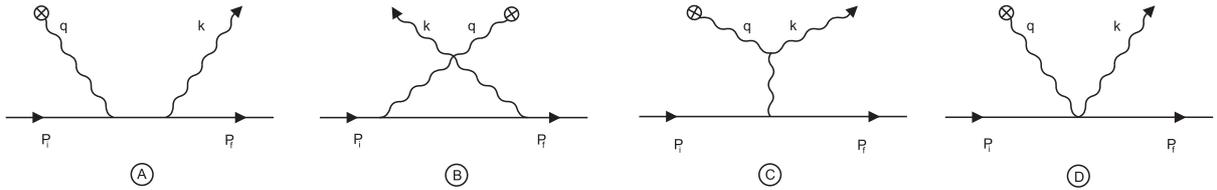}
\caption{Lowest order Feynman diagrams}
\label{fig1}
\end{center}
\end{figure}

We consider a particle of mass $m$, described by a non-hermitian scalar
field $\fii(x)$, which is scattered by a fixed central gravitational
potential induced by a mass $M$. The $\fii$-particle looses
energy-momentum and emits a (massive or massless) graviton. The
interaction with the classical central potential $\phi$ will be treated
according to the external field approximation where in the Newtonian
limit, $\phi$ becomes the Newtonian potential $\phi_N$. In lowest order,
the process is described by four second-order Feynman diagrams according to (Fig.\ref{fig1})
where $p_i^\mu=(E_i,\vec p_i)$, $p_f^\mu=(E_f,\vec p_f)$ are
the four-momenta of the initial and final particle, $k^\mu=(\omega,
\vec k)$ is the momentum of the bremsstrahl graviton, with $k_\mu
k^\mu=m_0^2$. We will assume the mass $m$ of the particle to be much
smaller than the central mass $M\gg m$ and the scattering takes place
under a small angle $\Theta'\ll m/E_i$ which in the classical picture
corresponds to a large impact parameter $\ro$. Then we are able to use
the first Born approximation for the external field $\phi_N$. Since at
the end, we want to compare our results with the classical theory, we
consider the case of ultra-relativistic particle energy ($m\ll E_i$) and
low frequency ($\omega\ll E_i$). This limit is known as the only
overlapping region of the Born and quasiclassical approximation.

According to our findings in the last section, we consider the S-matrix
element for each massive graviton mode $\psi_a,\,a=1,\ldots,6$ in the 
final state separately. Again, we choose the coordinate system such that
the outgoing graviton has momentum $k_\mu=(\omega,0,0,k_3)$. In the
scalar product $(\phi_f,S\phi_i)$, where $\phi_i$, $\phi_f$ is the initial, final state, respectively, only 
the emission operators $a_{\mu\nu}^+$ 
and $d^+$ in the final bremsstrahl graviton state are to be contracted 
with the coupling term \ref{coup} because there is no coupling to $v^\mu$.
In $\psi_2,\ldots,\psi_5$, these contributions are of order $O(m_0)$. Consequently,
the matrix elements of these final states are of order $O(m_0)$, too, compared to the
contributions $O(1)$ of $\psi_1$ and $\psi_6$. This proves that in the
massless limit $m_0\to 0$, the cross sections converge to the massless
ones. 

From now on, we consider bremsstrahlung of massless gravitons only. The
S-matrix element has the form
\begin{equation}
(\phi^{\ro\si}_f,S_A\phi_i)=\delta(E_f+E_k-E_i)M_A^{\ro\si}
\end{equation}
for each Feynman diagram A,B,C,D. Note that we have only energy
conservation. Momentum is not conserved because the central big mass $M$
absorbs a tiny amount of momentum, but its repulsion is neglected. The
external Newtonian potential is in momentum space given by
\begin{equation}
\phi_N(\vec q)=-\sqrt{\frac{2}{\pi}}\frac{G_NMm}{\vec q^2}
\end{equation}
where $G_N$ is Newton's constant. For a general final graviton state $a^{\ro\si+}\Omega$, we then find the following matrix element for diagram A:
\begin{eqnarray}
M^{\ro\si}_{A}&=&\frac{i\kappa^2\pi^{-5/2}}{8\sqrt{E_iE_f\omega}}\,\frac{G_NMm
}{\big|\vec p_f+\vec k-\vec p_i\big|^2}\nonumber\\
& &\cdot\frac{\big(E_i(E_f+\omega)-\frac{m^2}{2}\big)}{m^2-(p_f+k)^2-i0}\,p_{f\mu}
(p_f+k)_\nu\,t^{\ro\si\mu\nu}
\end{eqnarray}
where the coupling constant $\kappa$ is given in terms of Newton's
constant by (\cite{GHOST})
\begin{equation}
\kappa^2=32\pi G_N
\end{equation}
The result for diagram B is very similar. The final matrix elements are
\begin{eqnarray}
M^{\ro\si}_{A,B}&=&\frac{i\kappa^2\pi^{-5/2}}{8\sqrt{E_iE_f\omega}}\frac{G_NMm}{\big|\vec p_f+\vec k-\vec p_i\big|^2}\frac{E_{i,f}(E_{f,i}\pm\omega)
-\frac{m^2}{2})}{m^2-(p_{f,i}\pm k)^2-i0}\nonumber\\
& &\cdot\Big(\frac{1}{2}\big[p_{f,i}^\ro(p_{f,i}\pm k)^\si +p_{f,i}^\si(p_{f,i} 
\pm k)^\ro\big]-\frac{1}{4}p_{f,i}(p_{f,i}\pm k)\,\eta^{\ro\si}\Big)
\end{eqnarray}
where the upper sign and first subscript $f$ is for diagram A and the
other for B. For diagram C we obtain
\begin{eqnarray}
M^{\ro\si}_{C}&=&\frac{i\kappa^2\pi^{-5/2}}{8\sqrt{E_iE_f\omega}}\frac{G_NMm}{\big|\vec p_f+\vec k-\vec p_i\big|^2}\Bigg(-\frac{\omega(E_f-E_i)}{2(p_{f}-p_i)^2-i0}\big[p_f^\ro p_i^\si+p_i^\ro p_f^\si-2p_fp_i 
\,\eta^{\ro\si}\big]\nonumber\\
& &-\frac{1}{2(p_{f}-p_{i})^2-i0}\big[(p_i-p_f)^\si\eta^{0\ro}+(p_{i}-p_{f} 
)^\ro\eta^{0\si}-\frac{1}{2}(E_i-E_f)\eta^{\ro\si}\big]\nonumber\\ 
& &\;\;\;\cdot\,\big[(E_fp_i+E_ip_f)k-\omega p_fp_i+m^2\omega\big]\Bigg)
\end{eqnarray}
and diagram D gives
\begin{equation}
M^{\ro\si}_{D}=\frac{i\kappa^2m^2\pi^{-5/2}}{32\sqrt{E_iE_f\omega}}\frac{G_NMm}{\big|\vec p_f+\vec k-\vec p_i\big|^2}\Big(\eta^{\ro 0}\eta^{\si 0}-
\frac{1}{4}\eta^{\ro\si}\Big)
\end{equation}
By estimating the contributions from the 4 diagrams, we find that C is of order $O(\omega^2/m^2)$ and D of order $O(\omega m^2/E_i^3)$
compared to $O(1)$ for A and B. Therefore, restricting ourselves to the
low frequency region $\omega\ll E_i$, we neglect the contributions from C and D. The two
massless physical helicity states can be represented by the complex
polarization "tensor" $\eps_{\pm}^{\ro\si}(\vec k)$ introduced by Weinberg
\cite{WE}. We choose a coordinate system where the direction of the initial
momentum $\vec p_i$ is along the $z$-axis and the scattering takes place
in the $xz$-plane. Let $\Theta$ be the angle between $\vec p_i$ and
$\vec k$, $\Theta_{fk}$ the angle between $\vec p_f$ and $\vec k$, and $\Theta'$ is the scattering angle of the particle. Then the total matrix
element becomes (under the same assumptions as \cite{G})
\begin{equation}
M_\pm\approx\frac{2iG_N^2Mm}{\pi^{3/2}\omega^{3/2}E_i^2\Theta'^2}\eps_\pm
^{\ro\si*}(\vec k)\Big(\frac{p_{i\ro}(p_i-k)_\si}{1-v\cos\Theta}-
\frac{p_{f\ro}(p_f+k)_\si}{1-v\cos\Theta_{fk}}\Big)
\end{equation}
where $v=|\vec p_i|/E_i$ and the star is complex conjugation.
By squaring, summing over the two polarizations, and integrating over the
spherical angles of the emitted graviton, we find the cross section
\begin{equation}
\frac{d\si}{d\omega d\Theta'}\approx \frac{64\pi G_N^4M^2m^2E_i^2}{3\omega\Theta'}
\end{equation}
In order to make contact to the classical theory, we use the relation
between the scattering angle $\Theta'$ and the impact parameter $\ro$ in
the presence of a Newtonian field. In the ultra-relativistic case, it has
the form $\Theta'\approx 4G_NM/\ro$ and we use the definition of the
energy cross section
\begin{equation}
\frac{d{\cal E}}{d\omega}=\frac{\omega}{2\pi\ro}\frac{d\si}{d\ro
d\omega}
\end{equation}
With $\gamma=E_i/m$ we arrive at
\begin{equation}\label{dE}
\frac{d{\cal E}}{d\omega}\approx \frac{32G_N^4M^2m^4\gamma^2}{3\ro^2}
\end{equation}
In order to obtain the total radiated energy
\begin{equation}\label{E}
{\cal E}=\int\limits_0^{\omega_{th}}\frac{d{\cal E}}{d\omega}\,d\omega
\end{equation}
we must have an estimate for the threshold frequency $\omega_{th}$ of the low-frequency
region in the case of gravitational interaction. In order to compare with classical results, we insert the classical value (\cite{KT}) $\omega_{th}\approx\gamma/\ro$ in \ref{E}, and find the total energy radiated in a single
encounter
\begin{equation}\label{Etot}
{\cal E}^{\rm tot}\approx \frac{32G^4_NM^2m^4\gamma^3}{3\ro^3}
\end{equation}
Our result \ref{dE} agrees with the result obtained by Galtsov et al. \cite{G}
up to a factor of two. This factor is due to our two polarizations,
because Galtsov et al. have only treated scalar bremsstrahlung. There
have been studies on classical and semi-classical bremsstrahlung in the
sixties and seventies (\cite{KT} and references given therein). The classical 
picture mainly consists of two
massive colliding bodies, deflecting their trajectories and therefore
emitting gravitational waves. All those classical calculations arrived
at the total radiated energy ${\cal E}\sim\gamma^3/\ro^3$ in agreement with our result \ref{Etot}.

One important point to mention is the range of validity of the results obtained
from the Born approximation in quantum theory. Since the external field
is expanded in powers of $G_NMm$, we must have $G_NMm\ll 1$. According
to the discussion in \cite{KT} there are no astrophysical objects satisfying
this condition. Therefore, for the astrophysical situation some
resummation procedure is needed.

One might think that the massless limit of our massive gravity always agrees with general relativity. This is not the case. A counter-example is graviton-graviton scattering. In this process, the vector graviton states contribute. This is a consequence of the coupling term $4h^{\mu\nu}\d_\mu v^{\alpha}\d_\nu v_{\alpha}$ in \ref{T1} which survives in the limit $m_0\to 0$!

\end{document}